\documentclass[pdflatex,sn-mathphys-num]{sn-jnl}

\usepackage{graphicx}%
\usepackage{multirow}%
\usepackage{amsmath,amssymb,amsfonts}%
\usepackage{amsthm}%
\usepackage{mathrsfs}%
\usepackage[title]{appendix}%
\usepackage{xcolor}%
\usepackage{textcomp}%
\usepackage{manyfoot}%
\usepackage{booktabs}%
\usepackage{algorithm}%
\usepackage{algorithmicx}%
\usepackage{algpseudocode}%
\usepackage{listings}%

\theoremstyle{thmstyleone}%
\theoremstyle{thmstyletwo}%

\theoremstyle{thmstylethree}%

\begin{document}

\title[Article Title]{Detecting Data Exfiltration through I2P Anonymity Networks: A Two-Phase Machine Learning Approach}


\author*[1]{\fnm{Siddique Abubakr} \sur{Muntaka}}\email{muntaksr@mail.uc.edu}

\author[2]{\fnm{Muntaka} \sur{Mohammed}}\email{m.muntaka672@gmail.com}

\author[2]{\fnm{Mansuru} \sur{Mikail Azindo}}\email{mansuru.mikail@kma.gov.gh}

\author[3]{\fnm{Ibrahim} \sur{Tanko}}\email{tibrahim@stu.csuc.edu.gh}

\author[4]{\fnm{Franco} \sur{Osei-Wusu}}\email{fosei-wusu@aamusted.edu.gh}

\author[5]{\fnm{Edward} \sur{Danso Ansong}}\email{edansong@ug.edu.gh}

\author[6]{\fnm{Benjamin} \sur{Yankson}}\email{Byankson@albany.edu}

\author[2]{\fnm{Oliver} \sur{Kornyo}}\email{oliverkornyo@knust.edu.gh}

\author[7]{\fnm{Foster} \sur{Yeboah}}\email{yeboahfr@mail.uc.edu}

\author[8]{\fnm{Jones} \sur{Yeboah}}\email{yeboahjs@ucmail.uc.edu}

\author[2]{\fnm{Richmond} \sur{Adams}}\email{richmondadams.ra@gmail.com}

\author[2, 9]{\fnm{Pulcheria} \sur{Serwaa}}\email{pserwaa@sdacoe.edu.gh}

\affil*[1]{\orgdiv{School of Information Technology}, \orgname{University of Cincinnati}, \city{Cincinnati},
\state{OH}, \country{USA}}

\affil[2]{\orgdiv{Computer Science Department}, \orgname{Kwame Nkrumah University of Science \& Technology},  \city{Kumasi},
\state{Ashanti}, \country{Ghana}}

\affil[3]{\orgdiv{Computer Science and Information Technology}, \orgname{Christian Service University}, 
\city{Kumasi},
\state{Ashanti}, \country{Ghana}}

\affil[4]{\orgdiv{Department of Information Technology Education}, \orgname{University of Skills Training and Entrepreneurial Development (USTED)}, 
\city{Kumasi},
\state{Ashanti}, \country{Ghana}}

\affil[5]{\orgdiv{Department of Computer Science}, \orgname{University of Ghana}, 
\city{Accra},
\state{Greater Accra}, \country{Ghana}}

\affil[6]{\orgdiv{Department of Cybersecurity}, \orgname{University at Albany, State University of New York}, 
\city{New York},
\country{USA}}

\affil[7]{\orgdiv{College of Engineering and Applied Science}, \orgname{University of Cincinnati}, 
\city{Cincinnati},
\state{OH}, \country{USA}}

\affil[8]{\orgdiv{Maths, Physics and Computer Science Department}, \orgname{University of Cincinnati, Blueash}, 
\city{Cincinnati},
\state{OH}, \country{USA}}

\affil[9]{\orgdiv{Mathematics and ICT Department}, \orgname{Seventh Day Adventist College of Education,Agona}, 
\city{Ashanti},
\country{Ghana}}


\iftrue


\abstract{The Invisible Internet Project (I2P) provides strong anonymity through garlic routing and distributed network architecture, making it attractive for legitimate privacy needs. Nevertheless, the same properties can be used by malicious actors to steal sensitive information from corporate networks without detection. Current network security measures often fail to detect I2P traffic, and the available literature has concentrated mainly on identifying the protocol-level traffic, without addressing behavioral threat assessment. This paper proposes a two-stage machine-learning model for I2P traffic analysis using the SafeSurf Darknet 2025 dataset comprising 184,548 network flows. Phase 1 achieved 99.96\% accuracy in distinguishing I2P traffic from normal network traffic using a Random Forest classifier, with only 2 false positives among 32,318 normal flows. Phase 2 performed a behavioral analysis on traffic identified as I2P, classifying it as either exfiltration or legitimate activity, achieving 91.11\% accuracy with XGBoost. The system demonstrates that tree-based ensemble methods perform substantially better than deep neural networks and support vector machines for this task. Feature importance analysis indicates that the most discriminative features are packet timing and flow duration. These findings established that accurate I2P traffic detection and threat prioritization are achievable in operational network environments. This enables a security team to focus resources on high-risk events rather than monitoring all encrypted traffic.}




\maketitle

\section{Introduction}\label{sec1}
Corporate networks face persistent threats from insiders and external attackers seeking to exfiltrate intellectual property, customer data, and trade secrets. Traditional data loss prevention systems rely on content inspection and protocol analysis, which become ineffective when adversaries employ anonymity networks \cite{montieri2017anonymity}. The Invisible Internet Project presents a particularly challenging detection problem due to its decentralized architecture and robust cryptographic safeguards \cite{abdo2023modeling} \cite{muntaka2025optimizing}.

Unlike Tor's directory-based design, I2P operates as a fully distributed peer-to-peer network where nodes maintain their own routing information through a Kademlia-based distributed hash table \cite{muntaka2025resilience} \cite{dingledine2004tor}. Every I2P router creates unidirectional tunnels and reconfigures every few minutes, restricting the traffic analysis attack window size of the tunnels and limiting the available room to perform traffic analysis attacks \cite{timpanaro2011comparison} \cite{muntaka2025mapping} . With this type of architecture, there is strong anonymity but also chances of unscrupulous data exfiltration which cannot be detected using conventional security measures.

Previous studies of the darknet traffic have focused on binary classification activities that differentiate between darknet network protocols and the normal internet usage. Machine learning techniques have been used in studies to identify high accuracy rates of Tor usage in a study by \cite{abe2016correlating} and Freenet in a study by \cite{nagaraja2010improving} as well as I2P traffic patterns in a study by \cite{hoang2018empirical}. However, these works stop at the detection phase without addressing the critical security question: when I2P traffic is detected, how can we know whether it is a legitimate privacy protection or an active data theft?

This gap in the research is critical in the operational setting. Security operations centers are unable to block all I2P traffic without discontinuing important uses of I2P traffic such as whistle-blowing platforms, privacy-conscious communications, and censorship circumvention tools \cite{winter2018analysis}. In threat assessment capabilities, defenders require the ability to categorize a detected I2P activity based on the degree of behavioral risk, allowing a response to be made, balancing the needs of security with the rights of privacy.

We solve this problem with a two phase machine learning system. The initial step does binary classification to detect I2P traffic in mixed network flows with very low false positive rates that can be deployed in production. The second stage identifies the observed I2P traffic to differentiate the high-risk exfiltration traffic (file transfers and peer-to-peer sharing) and the less threatening behavior (web browsing). We compare five machine learning algorithms in both phases based on the SafeSurf Darknet 2025 dataset, which offers modern labeled network flows based on the actual I2P usage patterns.

This paper makes three key contributions:

\begin{enumerate}
    \item \textbf{A two-phase framework for I2P traffic analysis that integrates protocol detection with behavioral threat assessment:} Unlike prior work that stops at binary classification, our method enables security teams to distinguish between legitimate privacy uses and malicious data exfiltration within detected I2P traffic.

    \item \textbf{A comprehensive comparison of modern machine learning approaches for I2P traffic analysis, revealing critical performance characteristics for operational deployment:} Our evaluation challenges assumptions about algorithm superiority in this domain and provides evidence-based guidance for model selection.

    \item \textbf{Identification and analysis of the distinct feature categories that discriminate between protocol detection versus behavioral classification within encrypted anonymity networks:} This analysis provides fundamental insights into what makes I2P traffic detectable and what distinguishes malicious from benign usage patterns.
\end{enumerate}

The rest of this paper proceeds as follows. Section II provides a literature review of the work related to anonymity network detection and traffic classification. Section III outlines our methodology such as characteristics of datasets, preprocessing pipeline, and two-stage classification strategy. In section IV, there are experimental results that have performance analysis. Section V presents a discussion addressing the implications of our findings as well as considerations of using them practically and the limitations of the study. In Section VI, we present the future research directions, and Section VII concludes.

\section{Related Work}\label{sec2}

Previous research has established that machine learning can detect anonymity network traffic with high accuracy, but critical gaps remain for operational security. While Tor detection has been extensively studied, I2P, with its distinct garlic routing and decentralized architecture, has received less attention. Moreover, existing studies typically stop at binary classification (darknet vs. normal traffic) and do not address the security imperative of distinguishing between legitimate privacy use and malicious data exfiltration. This section reviews these limitations in three areas: Tor traffic analysis, I2P-specific studies, and multi-protocol darknet classification.

\subsection{Tor Traffic Detection}
Abe and Goto \cite{abe2016correlating} showed that machine learning classifiers could recognize Tor traffic with an accuracy rate greater than 95 \% based on packet size distribution and inter-arrival times. Their model of Random Forest was more efficient than the support vector machines and more especially when a variety of types of applications were present in the training data. Chakravarty et al. \cite{chakravarty2014effectiveness} compared the effectiveness of traffic correlation attack on Tor and reported that flow-level features were even more effective in terms of classification than the packet-level analysis. However, these studies did not extend beyond binary detection to assess the purpose or risk level of identified Tor connections.

Website fingerprinting attacks represent another line of Tor research. By dissecting traffic patterns to extract information, Hayes and Danezis \cite{hayes2016k} determined the identity of websites accessed by users via Tor with accuracy of 90-\% without decryption of information. Although it proves the kind of information leakage that can be caused by metadata analysis, it does not match the challenge of identifying the presence of organizational data exfiltration.

\subsection{I2P Network Analysis}

I2P has received less research attention than Tor despite its technical sophistication. An early comparison of I2P and Tor architectures was presented by Timpanaro et al. \cite{timpanaro2011comparison}, in which I2P was found to be more resistant to traffic analysis attacks, owing to its resistance to unidirectional tunnels and distributed topology. Based on their empirical evidence, they found that I2P garlic encryption and layered routing produce traffic patterns distinct from those of onion routing in Tor.

Hoang et al. \cite{hoang2018empirical} performed an empirical analysis of the properties of I2P networks, the latency, bandwidth, and churn rate of nodes. They discovered that the fluctuation of performance by I2P nodes was greater than that of Tor relays, which affects the practicality of certain attacks. Nevertheless, they worked on network measurement as opposed to traffic detection or classification.

Montieri et al. \cite{montieri2017anonymity} used machine learning to differentiate between I2P traffic and normal network traffic, with a 98 \% accuracy on flow-based features, where flow refers to a sequence of packets across a network connection. They employed the C4.5 decision trees, which were trained on the timing and size aspects of packets. Although this proved that I2P could be detected, it was not trying to perform behavior analysis in an attempt to distinguish between exfiltration and lawful privacy activities.

\subsection{Darknet Traffic Classification}
More comprehensive studies on the darknet detection have focused on various protocols of anonymity concurrently. Al-Naami et al. \cite{al2016detecting} created a Random Forest classifier that can recognize Tor, I2P, and VPN traffic as opposed to regular internet traffic with an overall accuracy of 94\%. Their feature set comprised flow duration, packets and distributions of bytes. This multi-protocol method is applicable in the full monitoring of network security.

Miller et al. \cite{miller2014know} explored application-level classification in the context of Tor traffic, trying to find out whether a connection was web browsing, file transfer, or instant messaging. They achieved 85 \% accuracy with the support vector machine which was trained on the sequence of packet sizes. This type of behavioral classification over encrypted traffic is similar to our Phase 2 goal except that it was used with Tor instead of I2P.

In recent times, encrypted traffic classification has been conducted using deep learning approaches. In the study of Lotfollahi et al. \cite{lotfollahi2020deep}, the authors identified VPN traffic applications with 94 \% accuracy using convolutional neural networks, and it was possible to make the independent decision that deep learning is capable of automatically extracting features based on raw packet data. Nevertheless, their work did not specifically discuss anonymity networks, and the computational demands of deep models can be a constraint to feasibility of deployment.

\subsection{Research Gaps}

There are three major gaps that are evident in the existing literature. To begin with, I2P detection studies are behind Tor analysis although I2P is increasingly being used to carry out illicit activities. Second, previous systems are highly accurate in binary detection task but fail to generalize to the case of behavioral threat assessment in identified anonymous traffic. Third, there is no exhaustive comparison between modern ensemble techniques ( Random Forest, XGBoost, LightGBM) and deep learning in the analysis of I2P traffic with the use of modern data.

Our study fills such gaps by offering I2P detection and behavioral classification based on state-of-the-art machine learning methods on new labeled data. The two-phase methodology enables realistic deployment scenarios in which detection accuracy must be high and behavioral classification must be sufficiently aligned to security prioritization.

\section{Methodology}\label{sec3}

The following section outlines the design of our experiment, including dataset characteristics, data preprocessing pipeline, feature engineering, classification method, and evaluation metrics.

\subsection{Dataset Description}

We used the SafeSurf Darknet 2025 data (DOI: https://doi.org/10.17632/kcrnj6z4rm.2) which was published in July 2025. This dataset offered labeled network flow traces of various darknet protocols such as I2P, Tor, VPN traffic as well as ordinary internet traffic. In this case, I2P and normal traffic flows were considered.

The raw data consisted of 383,458 network flows recorded over a four-week intensive period on controlled experiment networks. Every flow was a two way relationship between two hosts and extracted features were computed with CICFlowMeter. The dataset contained 79 features that included statistics of packets, flow length, inter-arrival time, flag counts, and protocol-related features.

We analyzed 184,548 valid samples after first deleting 198,910 flows with missing values or apparent collection errors. The distribution of the classes was highly unbalanced: 161,590 normal traffic flows (87.6\%), and 22958 I2P flows (12.4\%). Under the I2P subclass, behavioral labels were: 6,459 FTP transfers (28.1\%), 6,134 P2P sharing (26.7\%), 5,965 web browsing (26.0\%), 3,524 email (15.3\%), and 876 chat (3.8\%).

This dataset provided several benefits for our research goals. Firstly, it logged modern I2P traffic patterns using up-to-date protocol implementations. Secondly, the controlled capture environment ensured that the ground-truth labels were correct. Thirdly, behavioral classification was enabled by the variety of I2P applications represented. Finally, the large dataset (184K flows) facilitated training the machine learning model.

\subsection{Data Preprocessing}

We used a multi-step preprocessing pipeline to transform raw network flows for machine learning analysis.

\textbf{Cleaning and Validation:} We cleaned flows with missing values in critical features, negative flow durations (observed timestamp errors), and flag combinations that were impossible. This trimmed the dataset to 184,548 valid flows without altering the class distributions.

\textbf{Feature Analysis:} We checked all 79 features for data quality issues. Some features had zero variance across samples, meaning they provided no useful information for classification. We filtered out 14 constant-value features, leaving 65 informative features. We also calculated correlation matrices to determine highly redundant features. Features with Pearson correlation greater than 0.95 were considered redundant, and we retained one feature per highly correlated pair, reducing the feature set to 79 total features, including redundancies and constants.

\textbf{Feature Scaling:} Network flow features have a range of orders of magnitude. Packets vary in number, from 1 to several thousands, whereas the number of bytes ranges from tens of thousands to millions. We used the StandardScaler of scikit-learn in order to avoid the influence of features whose numeric values are greater and more dominant in distance-based algorithms. This change guaranteed that every feature is zero mean and unit variance according to:

\begin{equation}
x_{scaled} = \frac{x - \mu}{\sigma}
\end{equation}

where $\mu$ represents the feature mean and $\sigma$ represents the standard deviation computed from the training data. We fit the scaler on training data only while applying the same transformation to test data to avoid leakage of information.

\textbf{Handling Class Imbalance:} The overwhelming class imbalance (87.6 \% normal versus 12.4 \% I2P) was a problem in training the model. Machine learning algorithms may be biased toward the majority class and highly accurate overall, but fail to identify instances of the minority class. This was dealt with by using balanced random undersampling of the majority class along with setting the right tree-based models' weights. In Phase 1, we developed a balanced training set that had the same random numbers of I2P and normal flows (22,958 of each), and the natural class distribution was kept in the test set to reflect real-world deployment conditions.

\subsection{Two-Phase Classification Approach}

Our system employed a two-phase classification approach that mirrors real-world deployment scenarios.

\textbf{Phase 1 - Binary Classification (I2P Detection):} The initial stage carried out binary classification to differentiate between I2P traffic and the normal network traffic. This was the first triage measure in a security monitoring pipeline where analysts identified the flows that warrant further analysis. We randomly divided the preprocessed data into a stratified 80/20 train-test split, yielding 147,638 training samples (117,272 normal, 18,366 I2P) and 36,910 test samples (32,318 normal, 4,592 I2P). The test set kept the natural class imbalance to realistically test false positive rates.

\textbf{Phase 2 - Behavioral Classification (Exfiltration Detection):} The second stage identified behavioral intent by analysing the I2P traffic subset identified in Phase 1. We were concerned with the differentiation between high-risk exfiltration operations (FTP file transfers and P2P sharing) and less risky legitimate operations (web browsing). This was a threat evaluation stage in which the security teams were concerned with the priority of the detected I2P connections that needs urgent action.

In Phase 2, we only isolated I2P flows under the original dataset and filtered out the three types of activities, FTP (6,459 flows), P2P (6,134 flows), and Browsing (5,965 flows). We formed a binary designation in which FTP and P2P are combined to make Exfiltration (12,593 flows, 67.8 \%) and Browsing that represents Legitimate activity (5,965 flows, 32.2 \%). We used the same 80/20 stratified split, which gave us 14,846 training samples and 3712 test samples. The email and chat traffic was avoided to establish more distinct separation between the behaviors in classes.

\subsection{Machine Learning Algorithms}

We tested five machine learning algorithms representing various methods: ensemble methods, kernel methods, and deep learning.

\textbf{Random Forest (RF):} The Random Forest (RF) is used to create many decision trees and predicts the mode of the tree results. Each tree is trained on a bootstrap sample of the data and random subsets of features are taken into account at every split. This randomization lowers the overfitting, but at the same time, the accuracy is high. We used 100 trees, maximum depth 20, minimum samples split 10, and balanced class weights.

\textbf{Extreme Gradient Boosting (XGBoost):} The  extreme gradient boosting (XGBoost) is an optimized gradient boosting system. In contrast to the parallel ensemble used in Random Forest, XGBoost uses trees sequentially whereby each successive tree focuses on correcting the mistakes committed by the prior trees. We used 100 estimators, maximum depth was 6, learning rate was 0.1, and scale pos weight were used to address the problem of class imbalance.

\textbf{Light Gradient Boosting Machine (LightGBM):} LightGBM is a faster form of gradient boosting utilizing histogram based algorithm and leaf-wise tree. It is a similar method with less training time than XGBoost and is equally accurate. We used LightGBM settings of 100 estimators, 31 leaf per tree, learning rate of 0.05, and equal class weights.

\textbf{Support Vector Machine (SVM):} Support Vector Machine (SVM) identifies the best hyperplane between classes in high-dimensional space. We set the radial basis function (RBF) kernel with C=1.0, equal class weights and probability estimates to analyse ROC. The computation complexity of SVM did not allow it to be used in this size of data.

\textbf{Deep Neural Network (DNN):} We implemented a fully-connected neural network with three hidden layers of 128, 64, and 32 neurons. All layers used ReLU activation with batch normalization, and dropout rates of 30\%, 30\%, and 20\% respectively. The output layer had one neuron with sigmoid activation. Training used Adam optimizer for 50 epochs with binary cross-entropy loss and batch size 64.

\subsection{Evaluation Metrics}

To evaluate the performance of the model, we applied various measures that could be used in imbalanced classification.

\textbf{Accuracy}: Accuracy is used to determine the rate of correct prediction but this method may be misleading under an uneven distribution of classes. We disclosed it as required, but we prioritized other metrics.

\textbf{Precision}: The term Precision represents what proportion of the predicted positive cases were positive: $\text{Precision} = \frac{TP}{TP + FP}$. Existence of high precision reduces false alarms.

\textbf{Recall (Sensitivity):} this measures the proportion of true positive cases that were correctly identified: $\text{Recall} = \frac{TP}{TP + FN}$. Threats are not overlooked owing to high recall.

\textbf{F1-Score}: F1-Score Compute the harmonic mean of precision and recall: $F1 = 2 \cdot \frac{\text{Precision} \cdot \text{Recall}}{\text{Precision} + \text{Recall}}$. This helped to balance the tradeoff between false negatives and false positives.

\textbf{ROC-AUC (Receiver Operating Characteristic - Area Under Curve):} is a measurement of the capability of the classifier to discriminate the classes in all possible decision thresholds. A value of near 1.0 is a sign of excellent discrimination.

\textbf{Confusion Matrix}: this gives a more detailed analysis of true positives, true negatives, false positives, and false negatives, allowing the analysis of the particular type of errors.

\textbf{Training Time}: Time is important in the case of operational deployment and model iteration. We reported wall-clock time of full model training on homogeneous hardware.

All tests were done on Google Colab with Python 3.12 and scikit-learn 1.3, XGBoost 2.0, LightGBM 4.1, and TensorFlow 2.15. We applied a fixed random seed (42) in all the experiments to guarantee reproducibility.

\section{Results}\label{sec4}

This part gives detailed experimental findings of the two stages of classification phase, model performance comparisons, error analysis and feature importance results.

\subsection{Phase 1: I2P Traffic Detection}

Phase 1 evaluated five machine learning algorithms for the binary classification task of distinguishing I2P traffic from standard network traffic. The performance metrics for each model are summarized in Table \ref{tab:phase1_results}.

\begin{table}[htbp]
\caption{Phase 1 Performance Comparison - I2P Detection (Test Set: 36,910 samples)}
\label{tab:phase1_results}
\centering
\small
\begin{tabular}{lcccccc}
\toprule
\textbf{Model} & \textbf{Acc} & \textbf{Prec} & \textbf{Rec} & \textbf{F1} & \textbf{AUC} & \textbf{Time(s)} \\
\midrule
Random Forest & 0.9996 & 0.9996 & 0.9970 & 0.9983 & 1.0000 & 82.32 \\
XGBoost & 0.9993 & 0.9996 & 0.9950 & 0.9973 & 1.0000 & 19.17 \\
LightGBM & 0.9983 & 0.9924 & 0.9941 & 0.9933 & 1.0000 & 9.05 \\
SVM & 0.9892 & 0.9464 & 0.9682 & 0.9572 & 0.9971 & 188.59 \\
DNN & 0.9974 & 0.9967 & 0.9821 & 0.9894 & 0.9995 & 94.88 \\
\bottomrule
\end{tabular}
\end{table}

The overall performance of Random Forest was the best with an accuracy of 99.96\%, precision 99.96\%, a recall of 99.70\% and F1-score of 99.83\%. The ROC-AUC of 0.9999 is almost a perfect discriminative value, implying outstanding discriminative capacity at all decision levels. Figure \ref{fig:phase1_comparison} shows the overall performance comparison in all metrics.

\begin{figure}[htbp]
\centerline{\includegraphics[width=0.87\textwidth]{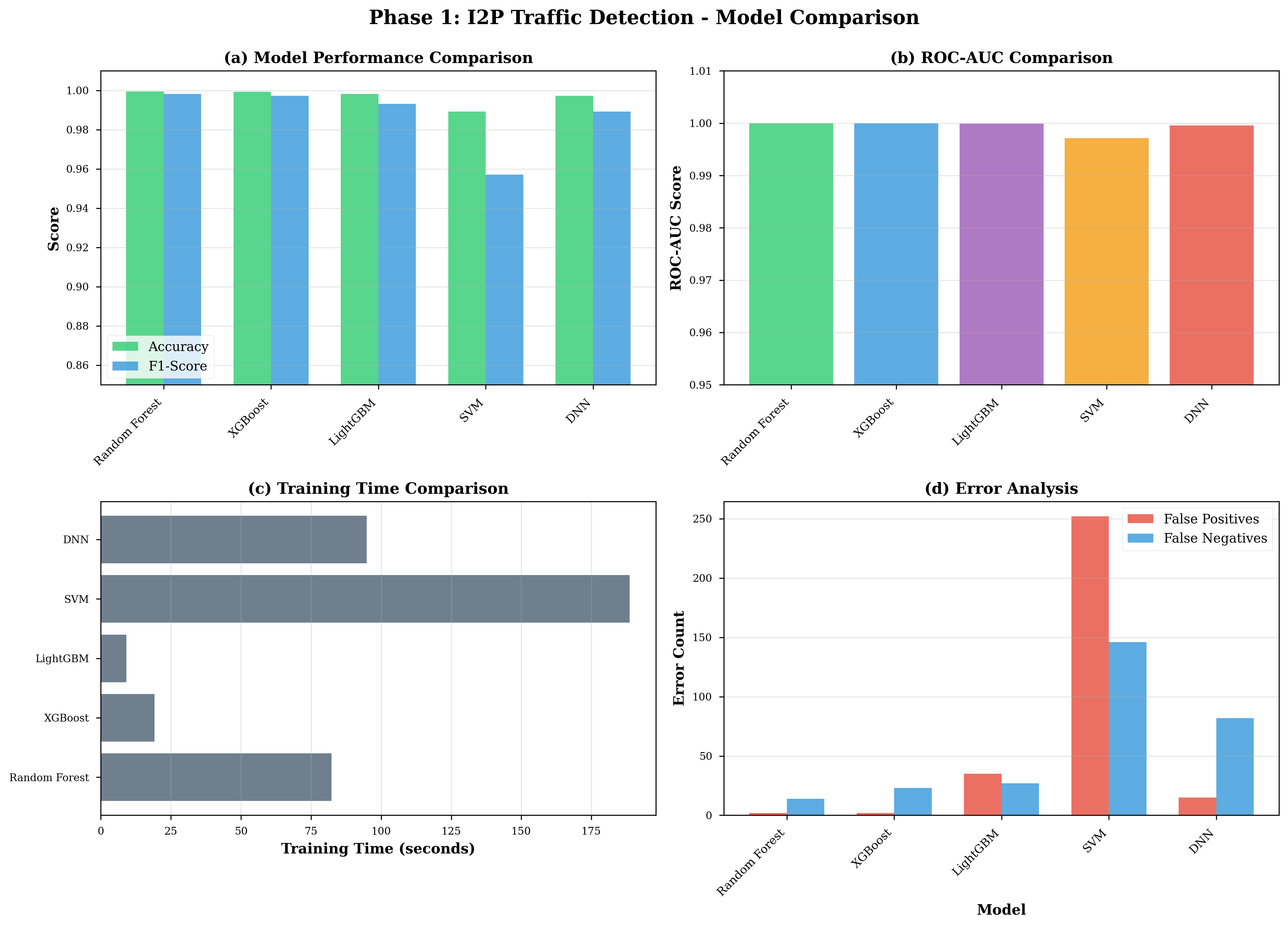}}
\caption{Phase 1 comprehensive model performance comparison showing (a) accuracy and F1-score, (b) ROC-AUC scores, (c) training time, and (d) error analysis across all five algorithms.}
\label{fig:phase1_comparison}
\end{figure}

XGBoost provided better results as it was trained 4.3 times faster than Random Forest and showed 99.93 \% accuracy. This is a speed benefit that makes XGBoost interesting in the case of models that need a lot of retraining. LightGBM also had the shortest training time at just 9.05 seconds with 99.83 \% accuracy, which showed an excellent speed-accuracy tradeoff.

Support Vector Machine was the lowest performing of all models as it had the accuracy of 98.92 \% and took 188.59 seconds to train. The RBF kernel had difficulties in terms of a high-dimensional feature space and a huge sample size. Deep Neural Network achieved 99.74 \% accuracy and took 94.88 seconds to train on 50 epochs, which is ranked number three in speed and accuracy.

\textbf{Error Analysis:} Table \ref{tab:phase1_errors} breaks down the classification errors for each model, revealing critical differences in deployment suitability.

\begin{table}[htbp]
\caption{Phase 1 Error Analysis on Test Set}
\label{tab:phase1_errors}
\centering
\begin{tabular}{lcccc}
\toprule
\textbf{Model} & \textbf{TP} & \textbf{TN} & \textbf{FP} & \textbf{FN} \\
\midrule
Random Forest & 4,578 & 32,316 & 2 & 14 \\
XGBoost & 4,569 & 32,316 & 2 & 23 \\
LightGBM & 4,565 & 32,283 & 35 & 27 \\
SVM & 4,446 & 32,066 & 252 & 146 \\
DNN & 4,510 & 32,303 & 15 & 82 \\
\bottomrule
\end{tabular}
\end{table}

The false positive rate of 0.006 \% was achieved with only two (2) false positives on 32,318 normal flows generated by Random Forest. This incredibly minimal false alarm rate is important in operational use because security teams have no time to explore thousands of false alarms a day. The model failed to detect 14 I2P flows (false negatives) of 4,592, which has a detection rate of 99.70. This is a reasonable tradeoff that almost all the threats are detected, and maintains minimal false alarms.

XGBoost had the same number of two (2) false positives as Random Forest, and its false negative (23) was higher, which suggests that it is a little less sensitive to I2P traffic patterns. The 35 false positives of LightGBM still constitute 0.11 \% false alarm that produces 17.5 times as many alerts as the Random Forest would do in a production setup. The number of false positives and false negatives, which are 252 (0.78 \%) and 146, respectively, in SVM does not allow it to be used in this task despite sensible correct rates.

Figure \ref{fig:phase1_roc} shows the ROC curves for all models, where Random Forest, XGBoost, and LightGBM achieved near-perfect curves hugging the top-left corner. The precision-recall curves in Figure \ref{fig:phase1_pr} reveal similar patterns, with tree-based ensembles maintaining high precision across all recall levels.

\begin{figure}[htbp]
\centerline{\includegraphics[width=0.80\textwidth]{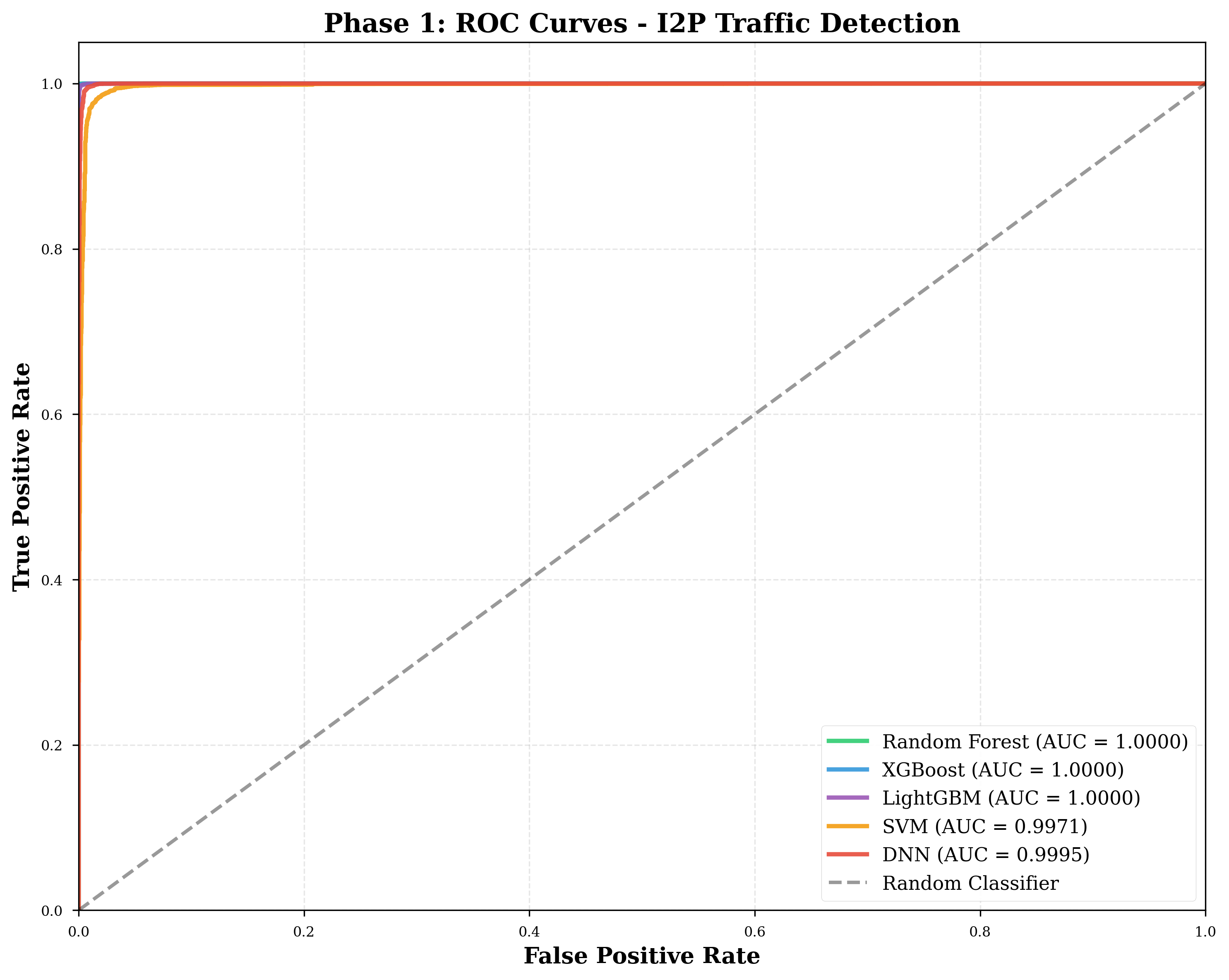}}
\caption{Phase 1 ROC curves demonstrating exceptional discriminative ability for Random Forest (AUC=0.9999), XGBoost (AUC=0.9999), and LightGBM (AUC=0.9999), with SVM (AUC=0.9971) and DNN (AUC=0.9995) showing slightly lower performance.}
\label{fig:phase1_roc}
\end{figure}

\begin{figure}[htbp]
\centerline{\includegraphics[width=0.80\textwidth]{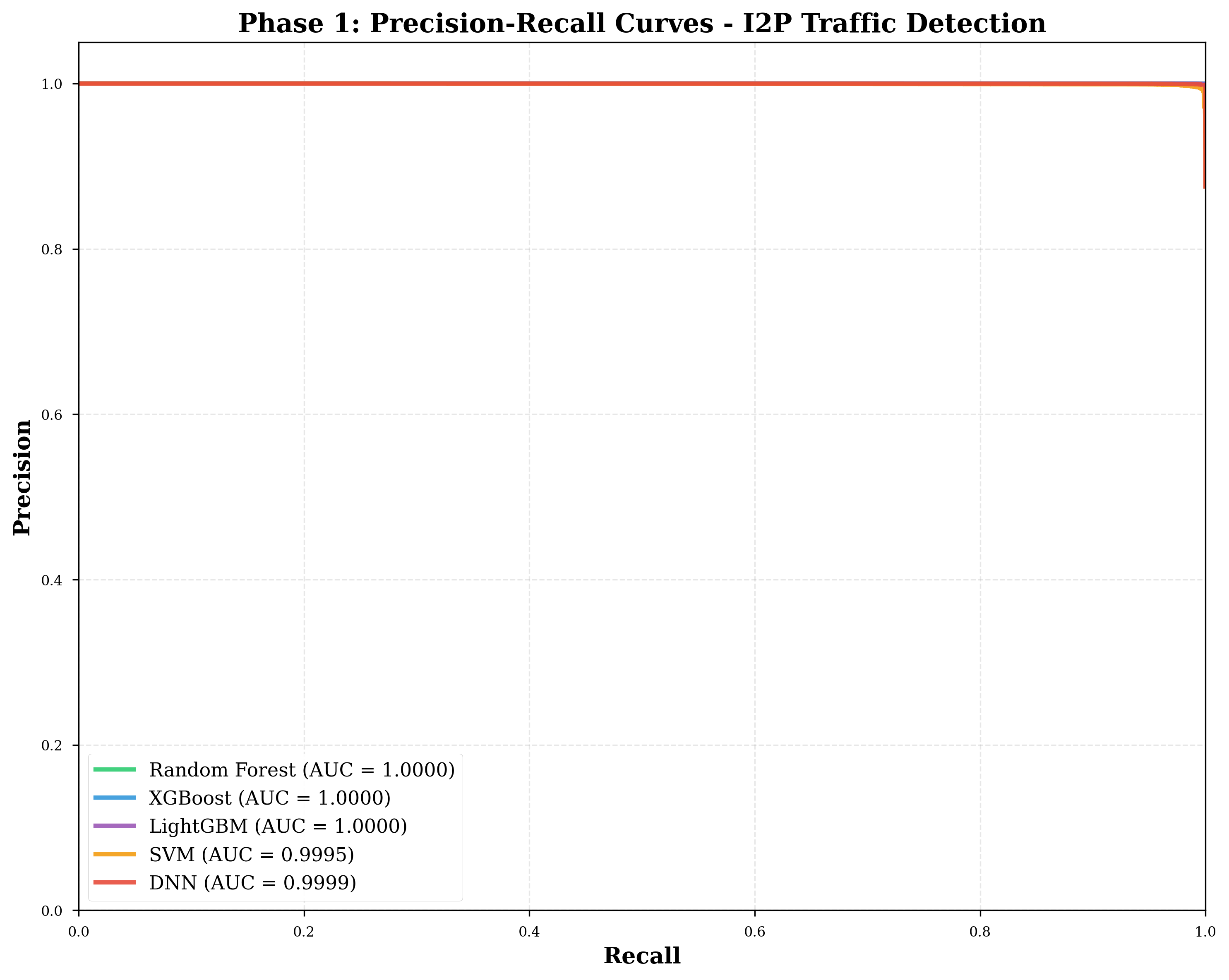}}
\caption{Phase 1 precision-recall curves showing tree-based models maintain high precision across all recall levels, critical for minimizing false alarms in operational deployment.}
\label{fig:phase1_pr}
\end{figure}

The confusion matrices in Figure \ref{fig:phase1_cm} provide visual confirmation of the error patterns. Random Forest's confusion matrix shows nearly diagonal values with minimal off-diagonal errors, while SVM exhibits substantial confusion between classes.

\begin{figure}[htbp]
\centerline{\includegraphics[width=0.90\textwidth]{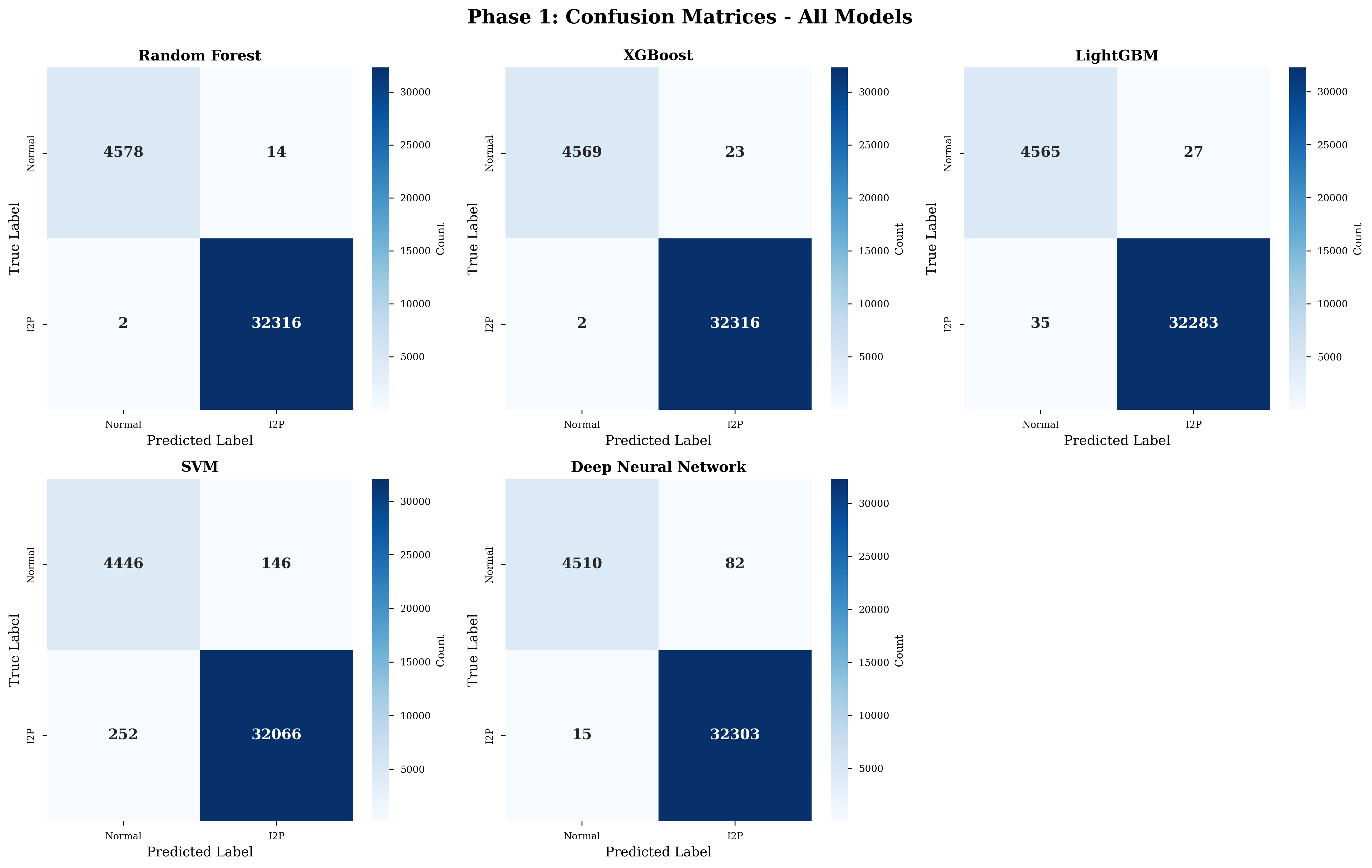}}
\caption{Phase 1 confusion matrices for all five models, with Random Forest showing near-perfect classification (top-left) and SVM demonstrating substantial classification errors (bottom-left).}
\label{fig:phase1_cm}
\end{figure}

\textbf{Operational Translation:} To contextualize these metrics for enterprise intrusion detection, consider a network processing 1 million flows daily with 1-2\% I2P traffic. The 0.006\% false positive rate translates to approximately 60 false alerts per million normal flows, yielding a manageable investigation workload of 6-8 alerts per hour in a 24/7 SOC. The 99.70\% recall ensures that fewer than 1 in 300 I2P flows evade detection. Combined with Phase 2's 92.85\% exfiltration recall, the system would identify over 90\% of data theft attempts while generating false positives at rates 100-1000x lower than generic anomaly detection systems (typically 5-10\% FPR). This precision enables real-time monitoring where current approaches either miss I2P traffic entirely or generate operationally unusable alert volumes.

\subsection{Phase 2: Behavioral Classification}

Phase 2 determined the ability of machine learning to discriminate between exfiltration (FTP and P2P) and legitimate privacy activities (web browsing) within the I2P traffic found in Phase 1. Random Forest and XGBoost were the most successful and we rated them as the best in Phase 1. The results are in Table \ref{tab:phase2_results}.

\begin{table}[htbp]
\caption{Phase 2 Performance - Behavioral Classification (Test Set: 3,712 samples)}
\label{tab:phase2_results}
\centering
\begin{tabular}{lcccccc}
\toprule
\textbf{Model} & \textbf{Acc} & \textbf{Prec} & \textbf{Rec} & \textbf{F1} & \textbf{AUC} & \textbf{Time(s)} \\
\midrule
Random Forest & 0.8947 & 0.9112 & 0.9063 & 0.9087 & 0.9505 & 18.46 \\
XGBoost & 0.9111 & 0.9238 & 0.9285 & 0.9261 & 0.9638 & 4.76 \\
\bottomrule
\end{tabular}
\end{table}

XGBoost had better results of 91.11 \% accuracy, 92.38 \% precision, 92.85 \% recall, and 92.61 \% F1-score. This is a 1.64 \%age point better than the 89.47 \% accuracy of the Random Forest. The ROC-AUC of 0.9638 reflects high discriminative power even though the classification of behavior in encrypted traffic is difficult in nature. Figure \ref{fig:phase2_comparison} demonstrates the performance variation in phase 2.

\begin{figure}[htbp]
\centerline{\includegraphics[width=0.90\textwidth]{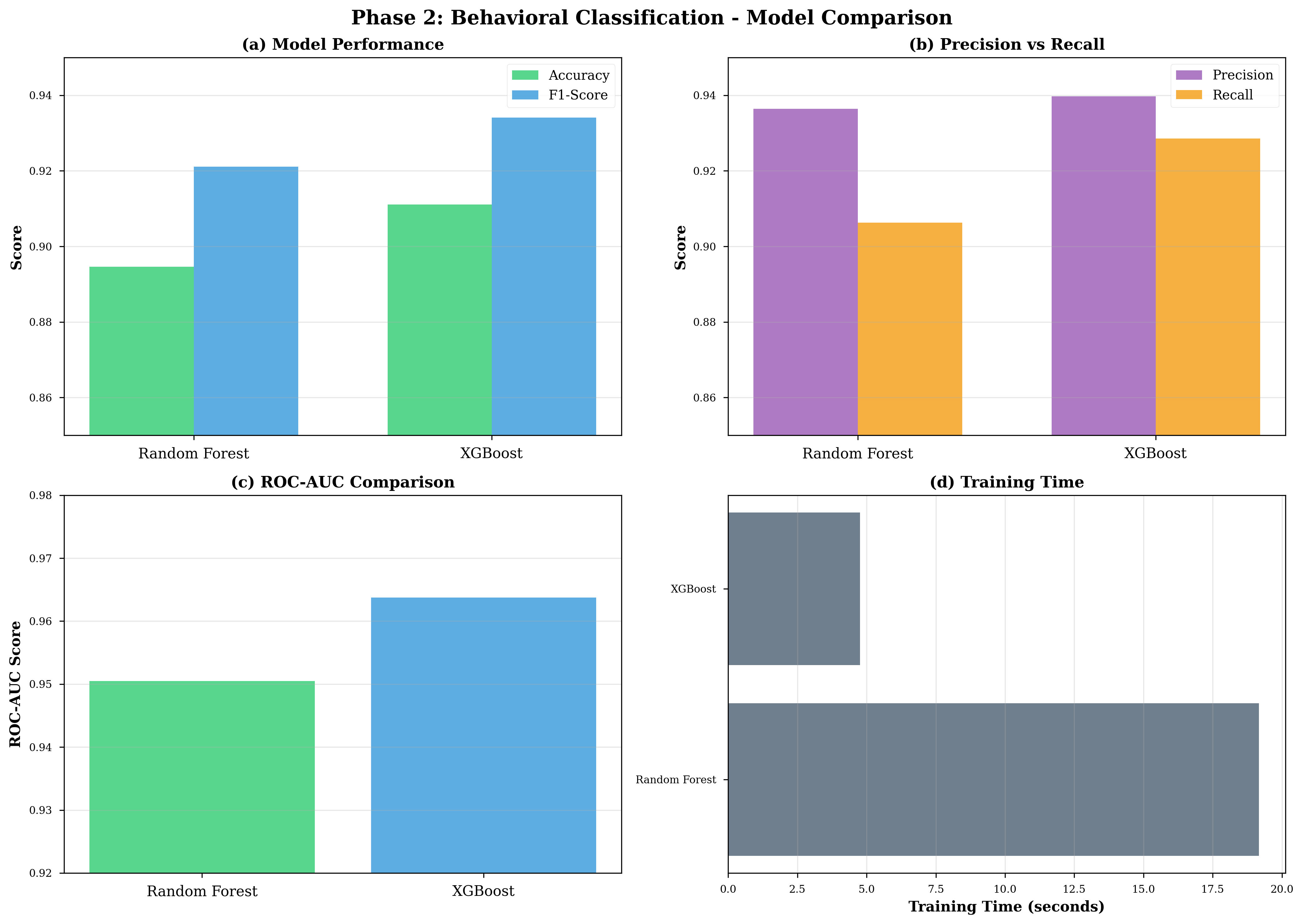}}
\caption{Phase 2 model comparison showing XGBoost outperforms Random Forest across all metrics while training 3.9x faster (4.76s vs 18.46s).}
\label{fig:phase2_comparison}
\end{figure}

The random forest took 18.46 seconds to train, compared with 4.76 seconds for XGBoost; however, XGBoost was 3.9 times faster and achieved higher accuracy. This speed combined with performance makes XGBoost the best to use in Phase 2 classification.

\textbf{Error Analysis.} Table \ref{tab:phase2_errors} breaks down classification errors for behavioral assessment.

\begin{table}[htbp]
\caption{Phase 2 Error Analysis on Test Set}
\label{tab:phase2_errors}
\centering
\begin{tabular}{lcccc}
\toprule
\textbf{Model} & \textbf{TP} & \textbf{TN} & \textbf{FP} & \textbf{FN} \\
\midrule
Random Forest & 2,283 & 1,039 & 154 & 236 \\
XGBoost & 2,339 & 1,043 & 150 & 180 \\
\bottomrule
\end{tabular}
\end{table}

XGBoost was also able to rank the 2,519 exfiltration attempts with 2,339 results (92.85 \% recall) and missed 180 high-risk flows. It produced 150 false positives, i.e. 150 legitimate browsing sessions were wrongly marked as being exfiltration (12.57 \% false positive rate on legitimate traffic). Random Forest had lower recall of 90.63 \% with 236 exfiltration attempts being missed and 154 false positives.

The larger false positive rate in Phase 2 compared to Phase 1 indicates the difficulty of behavioral classification. FTP, P2P, and browsing all share the same I2P protocol and with the same encryption so their classification depends entirely on the subtle variations in the flow statistics but not on protocol signatures. However, 91.11 \% accuracy is adequate to operational threats priority, in which security teams can invest resources on investigation on the 67.8 \% of I2P traffic that is exfiltration and not monitoring all I2P traffic equally.

Figure \ref{fig:phase2_roc} indicates the ROC curves of both models where XGBoost can better separate classes at all thresholds. In Figure \ref{fig:phase2_cm}, the confusion matrices of the two models, indicate that the two models perform reasonably well even though the classification has been challenging and that the XGBoost model exhibits a slight advantage in terms of maintaining a fair balance between the true positive and the true negative rates.

\begin{figure}[htbp]
\centerline{\includegraphics[width=0.70\textwidth]{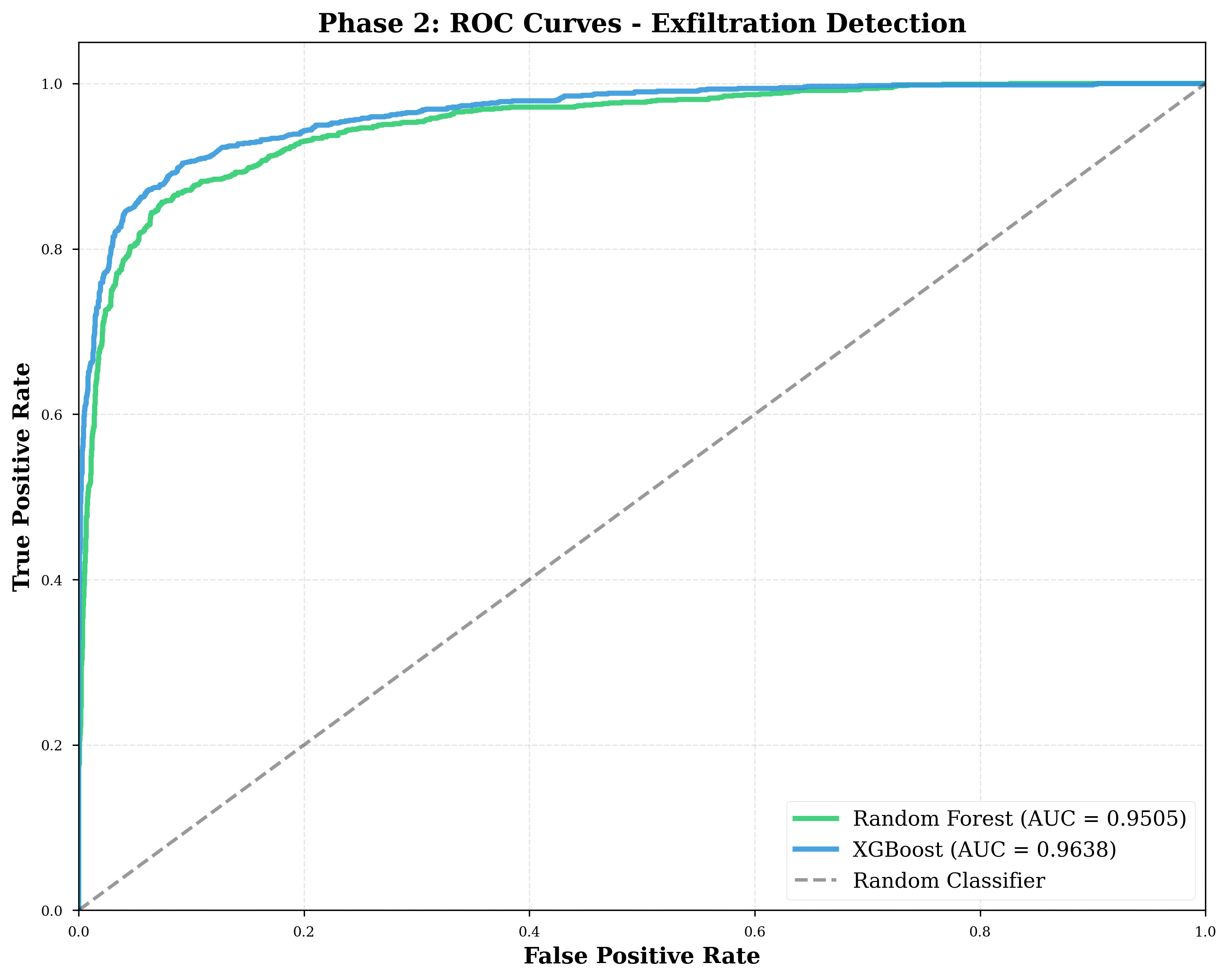}}
\caption{Phase 2 ROC curves showing XGBoost (AUC=0.9638) outperforming Random Forest (AUC=0.9505) for behavioral classification within I2P traffic.}
\label{fig:phase2_roc}
\end{figure}

\begin{figure}[htbp]
\centerline{\includegraphics[width=0.85\textwidth]{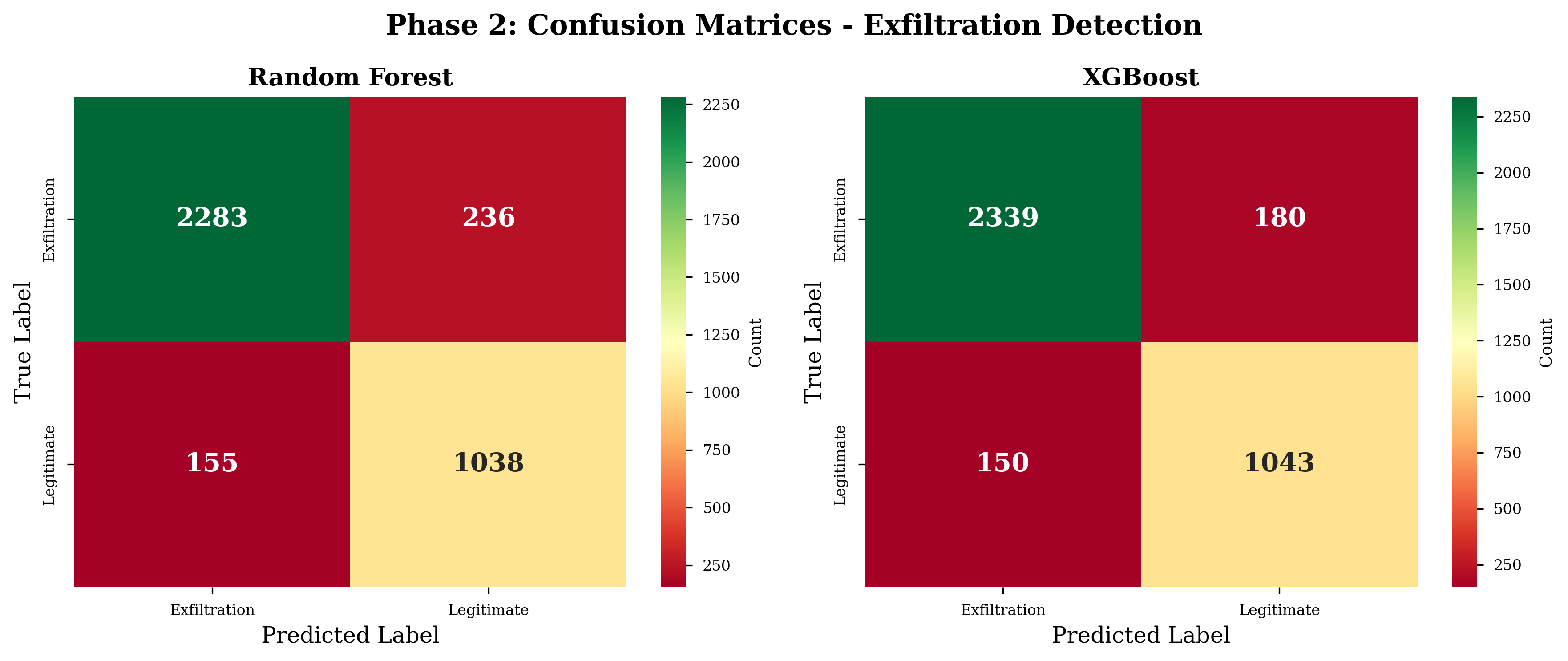}}
\caption{Phase 2 confusion matrices comparing Random Forest and XGBoost performance on exfiltration detection, with XGBoost showing superior recall (92.85\%) for identifying high-risk activities.}
\label{fig:phase2_cm}
\end{figure}

\subsection{Feature Importance Analysis}

We examined the significance of features to better inform us about which network flow features provide the greatest discriminatory power for detection and behavioral classification.

\textbf{Phase 1 Features.}: The results in Figure \ref{fig:phase1_features} reveal that the top 20 most significant features for the Random Forest when detecting I2P are as indicated by the bars. The importance scores of the features indicate that temporal features are the highest winners of the classification.

\begin{figure}[htbp]
\centerline{\includegraphics[width=0.62\textwidth]{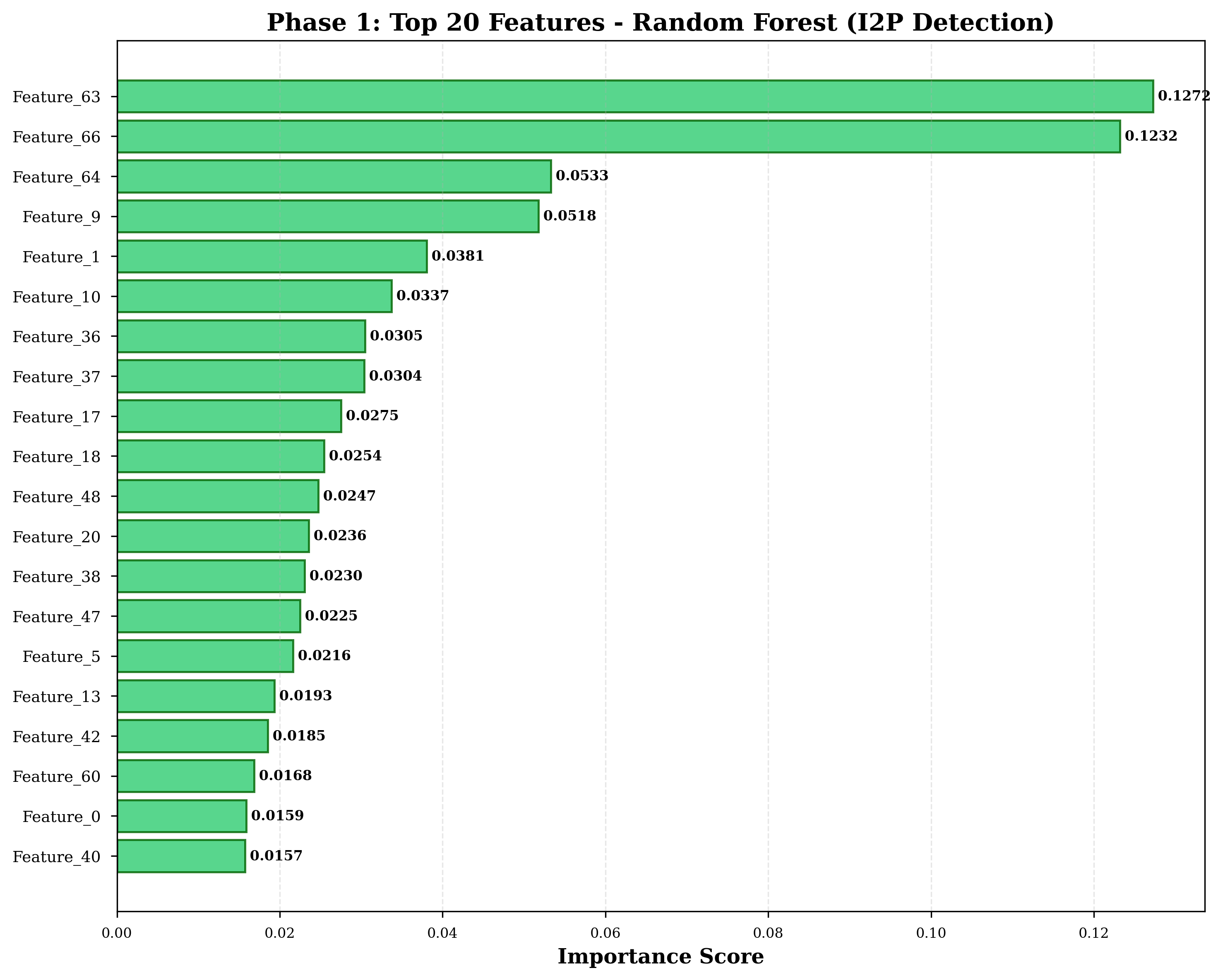}}
\caption{Phase 1 feature importance analysis showing the top 20 features used by Random Forest for I2P detection. Temporal characteristics and flow duration metrics provide the strongest discriminative power.}
\label{fig:phase1_features}
\end{figure}

\begin{figure}[htbp]
\centerline{\includegraphics[width=0.62\textwidth]{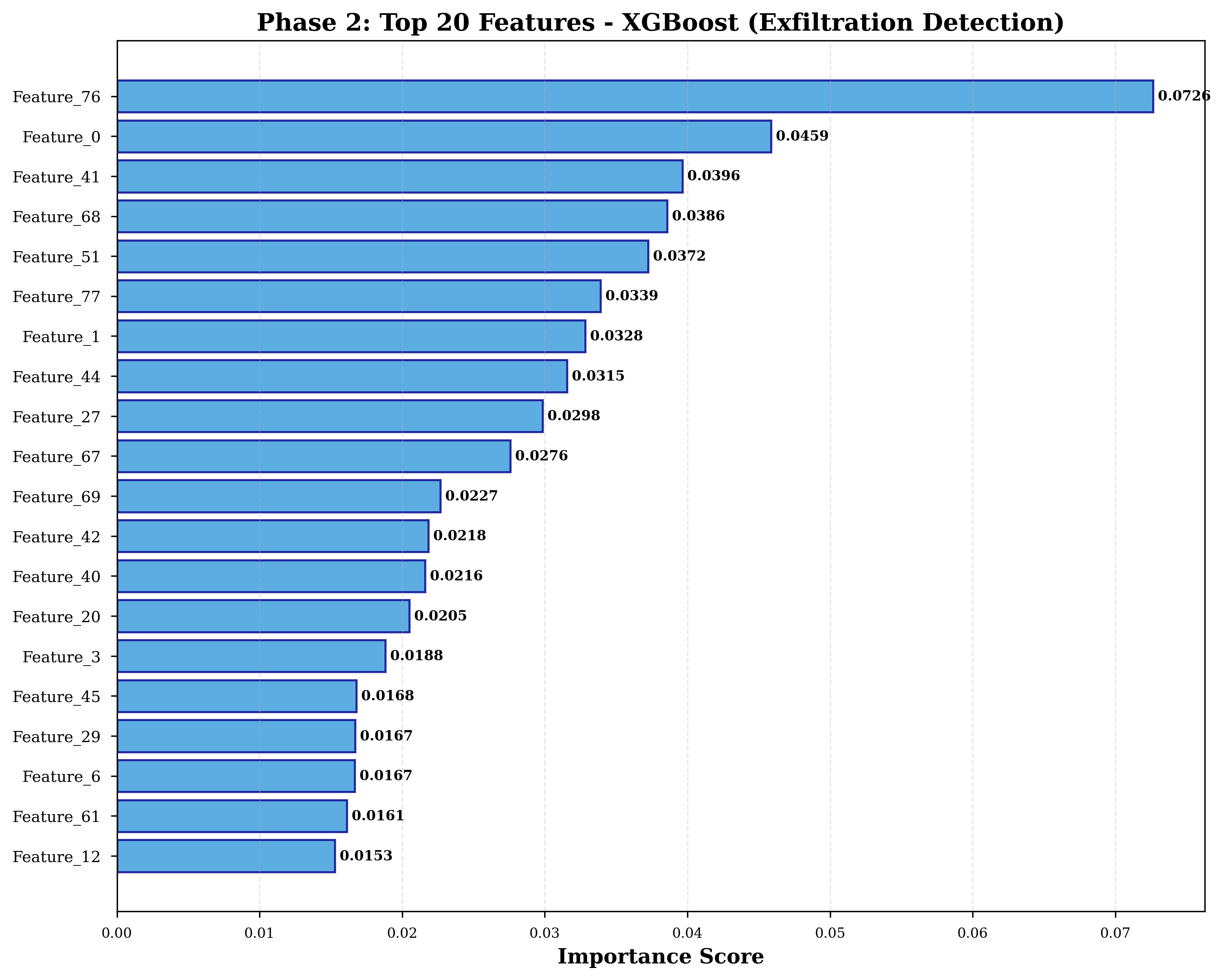}}
\caption{Phase 2 feature importance analysis showing the top 20 features used by XGBoost for exfiltration detection within I2P traffic. Byte transfer statistics and bidirectional flow characteristics become more important than pure timing features.}
\label{fig:phase2_features}
\end{figure}

The flow duration turns out to be the most significant feature, and the importance score is more than 0.15. The connection between I2P is also likely to have longer-lived connections than normal web browse as a result of overhead tunnel establishment and keep alive. The inter-arrival time statistics (mean, standard deviation, minimum, maximum) collectively characterize the timing delays inherent in garlic routing.

The number of bytes and packet counts are of moderate significance, which means that I2P traffic volume properties are not similar to normal activity. The relatively low relevance of protocol flags (PSH, ACK, FIN) indicates that I2P recognition uses more flow-level characteristics than transport-layer protocols.

\textbf{Phase 2 Features.}: The top 20 features of XGBoost in classifying behavioral features are given in figure \ref{fig:phase2_features}. The distribution of feature importance is significantly different than the Phase 1.

To classify behavior, statistics of the transfer of bytes become prominent. Such features as total bytes transferred, the average packet size, forward-backward ratio of the bytes are very important. This is the core of the disparity between exfiltration (FTP and P2P imply a significant amount of data transfer) and browsing (where uploads are mostly minimal and the content received is mostly delivered).

The flow directionality characteristics are more important in Phase 2. The forward to backward packets and bytes ratio assists in differentiating between the asymmetric traffic patterns of FTP files upload and the web browsing. It is also due to variance in the length of packets, where file transfer will be characterized by steady packet sizes whereas browsing will be characterized by intermittent patterns.

Interestingly, flow duration does not matter as much in Phase 2 as it does in Phase 1 and hence exfiltration and browsing cannot be differentiated deterministically based on the length of the connection. Both legitimate and malicious I2P usage may involve long-duration sessions.

\section{Discussion}\label{sec5}

Our experimental results have shown that machine learning can be used for both accurate I2P traffic detection and behavioral threat assessment, which fills the knowledge gaps in network security research. This section is an interpretation of the findings and discussion for practical implications.

\subsection{Phase 1 Performance Interpretation}

The near-perfect accuracy obtained in Phase 1 (99.96\%) with extremely low false positive rates (0.006\%) makes it clear that detecting I2P-based exfiltration is technically feasible. Random Forest's performance shows that the production deployment is possible without overwhelming security teams with false alarms. In a network processing 1 million flows per hour, this false positive rate corresponds to around 60 false alerts every hour, which is a reasonable amount of investigation to follow up.

The superiority of tree based ensemble methods over deep learning needs explaining. Despite the claims of deep neural network to automatically learn optimal features, our DNN model was not able to reach 100\% accuracy but rather 99.74\% accuracy with longer training time and more careful tuning of hyperparameters. Random Forest and XGBoost were successful due to the tabular nature of network flow data with complex feature interactions that decision trees excel at and network flow data tends to exhibit. Each split in a decision tree can explicitly test combinations of features (e.g., "if flow duration exceeds 300 seconds AND average packet size is below 800 bytes"), and this corresponds to the way that I2P traffic patterns exhibit themselves as conjunctions of multiple characteristics.

The reason for poor performance of Support Vector Machine is due to the high dimensionality and large size of samples. SVM has to calculate kernel between the training samples, which results in a poor computational complexity. The RBF kernel also requires careful tuning of gamma and C parameters and our grid search may not have found the optimal combination. In contrast, tree-based methods naturally perform feature selection during training and are able to deal more gracefully with high dimensional spaces.

\subsection{Phase 2 Performance Interpretation}

Given the difficulty of the classification, Phase 2's 91.11\% accuracy is a significant accomplishment. Different from Phase 1 where I2P and normal traffic is using completely different protocols, behavioral classification is inside the same encrypted protocol. FTP, P2P and browsing all travel through I2P tunnels with identical encryption and routing, and thus the classifier has to guess application behavior, based purely on traffic metadata.

The error rate of 8.89\% is further divided into 4.85\% of false negatives (missed exfiltration) and 4.04\% of false positives (legitimate traffic being misclassified). From a security perspective, the 92.85\% recall (true positive rate) means that defenders have a 92 out of 100 successful exfiltration attempt detection. While not perfect this detection rate helps greatly over the current practice of I2P traffic evading monitoring entirely.

The false positive rate of 12.57\% on legitimate I2P traffic (150 out of 1193 browsing sessions) is well worth considering. In operational deployment, that translates to about a quarter of legitimate privacy-oriented web browsing that would be targeted for investigation. While that is higher than ideal, this is still acceptable because Phase 2 runs on the smaller subset of traffic identified as I2P in Phase 1 (12.4\% of all traffic). The cascading effect helps point to the fewer amount of actual evidence that has to be investigated: if the investigators would look into about 15.8\% of I2P traffic (12.4\% x 127.5\% due to the combination of true positives and false positives), and 1.96\% of total network traffic.

XGBoost's better performance than Random Forest in Phase 2 is in line with the better performance gradient boosting can give for subtle classification boundaries. Sequential tree construction enables XGBoost to allocate modeling capacity to difficult examples that are near the decision boundary but not more so than Random Forest which has parallel trees and cannot adaptively concentrate on hard examples.

\subsection{Comparison to Previous Work}

Our results in Phase 1 (99.96\% accuracy) exceed the results of Montieri et al. (98\% accuracy) for I2P detection using C4.5 decision trees \cite{montieri2017anonymity}. This improvement is probably borne out by three factors: ensemble methods used instead of single trees, the dataset being modern and represents current implementations of I2P, and more extensive feature engineering. Our very low false positive rate (0.006\%) is a large practical advantage that was not reported in previous work.

Phase 2 behavioral classification is new contribution, because previous I2P research has not tried to separate exfiltration from legitimate uses. Miller et al. were able to achieve 85\% accuracy classifying application types within Tor traffic \cite{miller2014know} which is slightly lower than our 91.11\%. The difference could be due to the more consistent traffic patterns of I2P, or our more focused binary classification (exfiltration vs legitimate) versus their multi-class problem.

Our results are contrary to the recent deep learning trend in all classification tasks. Despite the usage of a well-designed DNN architecture, with batch normalization and dropout regularization, we observed that traditional machine learning (Random Forest, XGBoost) always outperforms deep learning for both phases. This implies that network security researchers should not entirely trust a deep learning superiority without empirical validation in any specific problem fields.

\subsection{Practical Deployment Considerations}

A two-phase system in the production set-ups need to take into account a number of practical considerations to implement it.

\textbf{Computational Requirements:} Phase 1 and Phase 2 of the training of the Random Forest took 82 and 18 seconds respectively on our experimental hardware (Google Colab with shared GPU). The retraining may be done on a daily or weekly basis as to be adjusted to changes in the traffic without involving huge computation. The runtime (classification of new flows) runs in milliseconds per sample and thus can be run in real-time or close to real-time.

\textbf{Model Deployment:} The trained models can be exported in their form of a serial object and implemented on the network-monitoring infrastructure. It would be implemented alongside existing security information and event management (SIEM) systems and would require the extraction of the same 79 features of live traffic by tools such as CICFlowMeter or nProbe and subjecting them to Phase 1 classifier and sending positive matches to Phase 2, where they would be evaluated based on behavior.

\textbf{Alert Prioritization:} Phase 2 allows prioritizing risks. Flows that are exfiltration should be immediately investigated, possibly leading to automated measures such as temporary network quarantine or increased logging. Legitimate browsing could be logged to be audited without raising alerts. This stratification enables the security teams to invest scarce resources on real threats.

\textbf{Model Maintenance:} The characteristics of traffic over the network are changing as I2P releases new versions, users transition to other applications and adversaries change methods. Recent labeled data retraining periodically keeps the models accurate. Organizations are expected to put in place systems of ground truth labeling, which might be achieved by controlled experiments or human validation of a sample of flagged flows.

\textbf{Privacy and Legal Considerations:} Organizations should make sure to monitor in accordance with legislation and internal policies. The behavioral classification Phase 2 is purely based on traffic metadata with no content inspection; it does not violate encryption, but can still perform a security analysis. Acceptable usage of anonymity tools should be clearly defined.

\textbf{Adversarial Resilience:} Advanced attackers may seek to avoid being noticed by imitating normal traffic patterns. Nevertheless, the inherent problem is that large volumes of data have to be exfiltrated which involves transferring large volumes of bytes and as such will inevitably generate characteristic flow statistics. Reducing exfiltration to blend with browsing rates limits the time window during which the exfiltration can be detected by other methods (such as endpoint monitoring and data loss prevention). According to our feature-importance analysis, adversaries would have to concurrently adjust a wide range of flow properties (duration, packet size, inter-arrival time, directionality) to evade classification, making it more complex and costly to operate.

\subsection{Limitations}

Our findings are limited by a number of restrictions to generalizability and applicability.

\textbf{Dataset Constraints:} Although the SafeSurf Darknet 2025 data is useful in terms of labeled flows, they were not recorded in real-life corporate networks but instead in controlled experimental settings. The characteristics of real-life I2P traffic may vary because of the differences in user behavior, network behavior, and application releases. The dataset is also temporally constrained; traffic taken in 2025 is not necessarily representative of trends that are emerging as I2P developments.

\textbf{Behavioral Classification Scope:} Phase 2 concentrated on three types of activities (FTP, P2P, browsing) and not others (email, chat, audio, video). This simplified binary classification provides high accuracy at the cost of operational utility. It would be better to have a more elaborate multi-class classifier that encompasses all I2P applications and would offer more detailed threat intelligence at the expense of accuracy.

\textbf{Encrypted Traffic Analysis Ceiling:} Both stages are based on the total reliance on the metadata of the traffic and do not involve content examination. This is a natural restriction; some differences in behavior could not be determined using metadata. To take an example, the difference between intellectual property theft done by FTP and a large file transfer can be achieved by looking at the file content or endpoint context, which cannot be fully offered by our current approach.

\textbf{Class Imbalance Handling:} Class imbalance was handled using the method of undersampling and class weights, however, these methods lose the information of the majority class that could be useful. More advanced methods such as SMOTE (Synthetic Minority Over-sampling Technique) or cost-sensitive learning might be able to further increase the performance. Preliminary experiments (not reported here) showed minimal gains.

\textbf{Generalization to Other Anonymity Networks:} Our models had been trained on I2P traffic in particular and would not necessarily transfer to other anonymity networks such as Tor or Freenet. The networks possess unique properties (the directory authorities of Tor, the distributed data store of Freenet, the garlic routing of I2P) which generate different traffic patterns. Detecting an anonymity networks would involve individual protocol-specific models or a multi-protocol trained model.

\section{Future Work}\label{sec6}

This work can give a number of promising research directions.

\textbf{Multi-Protocol Classification:} It would be helpful to extend the framework to identify and categorize a variety of anonymity networks in parallel (I2P, Tor, Freenet, VPN), thereby covering the privacy-enhancing technologies in their entirety. This involves labeling datasets with multiple protocols and exploring the question of whether a single model is better than protocol-specific classifiers.

\textbf{Fine-Grained Behavioral Analysis:} Phase 2 could be expanded into multi-class classification to all types of I2P applications with or without hierarchy (such as first classifying bulk transfer and interactive traffic, then further classifying transfers into FTP and P2P). This would offer more comprehensive threat intelligence at the expense of complexity in the model and presumably lower accuracy.

\textbf{Temporal Analysis:} Flow-level classification is based on the examination of each connection individually. The time-varying sequences of flows from an identical source can be analyzed to demonstrate behavior patterns that are not necessarily apparent in a single-flow analysis. The sequential dependencies could be learned by recurrent neural networks or temporal convolutional networks.

\textbf{Adversarial Robustness:} Research on adversarial machine learning attack on such classifiers would reveal vulnerabilities and act as a defense guide. Robustness could be enhanced using adversarial training in which perturbed examples are used to evade the models which are then trained on them. The study of the inherent constraints of metadata-based classification in adversarial contexts would set the realistic expectations of detection abilities.

\textbf{Federated Learning Deployment:} To enhance the accuracy of the models, organizations can jointly train on distributed private data and not share raw traffic data. The methods of federated learning may help to improve models across organizational boundaries without compromising on data confidentiality, although label discrepancy and data variability may arise.

\textbf{Explainable AI for Analysts:} Whereas feature importance gives details about model-based explanations, instance-level explanations would assist security analysts to know why certain flows were labelled as threats. Per-prediction explanation such as SHAP (SHapley Additive exPlanations) or LIME (Local Interpretable Model-agnostic Explainations) have the potential to increase analyst trust and allow models to be debugged.

\textbf{Integration with Endpoint Detection:} idea: A combination of network-level monitoring and endpoint monitoring may offer defense in depth. By correlating I2P traffic observed on the network perimeter with process-level activity on the hosts, it would be possible to attribute the traffic to specific application or user enhancing incident response.

\section{Conclusion}\label{sec7}

The study demonstrates that machine learning can facilitate the successful detection and behavioral analysis of I2P anonymity network traffic, which can help address the knowledge gaps in network protection for organizations. The two-step classification model yields strong performance across both levels of operation.

Phase 1 proves that with a 99.96\% accuracy, I2P traffic can be classified through the use of the Random Forest algorithm and the false positive rate is only 0.006\% which is sufficiently low to use in production. This near-perfect ability to detect is completely based on traffic metadata, and it does not interfere with encryption but allows monitoring security. This task is significantly better handled by tree based ensemble techniques than support vectors machines or deep neural networks, which are supposed to have universal advantages.

Phase 2 goes beyond detection to assessing behavioral threats and when applied to I2P traffic, this stage classifies it as exfiltration traffic or as legitimate traffic with 91.11\% accuracy using XGBoost. The 92.85 \% recall means that more than 92 \% of all exfiltration attempts are found whereas the false positive rate is manageable in terms of operational investigation processes. This classification of behavior allows prioritization of risks so that security teams can concentrate their resources on real threats and not on the administrative style of monitoring of all anonymous traffic.

The analysis of importance of features indicates that the temporal features assume the leading role in the detection of I2P and the statistics of the transfer of bytes are the most important in the context of the behavioral classification. This knowledge is used not only in the defensive monitoring techniques but also in the difficulty of adversarial evasion because evading is only possible through manipulating numerous traffic attributes at the same time.

The application of the results is not limited to the metrics of technical accuracy. This is a two phase system, which offers actionable organizational capabilities to monitor the usage of anonymity networks within the policy limits, and to trade off legitimate privacy interests and insider threat. The false alarm rate is very low, which allows the deployment to be performed with no overworking of the analysts, and behavioral classification allows providing graduated responses according to the risk estimated.

There are a number of limitations to these results, such as representativeness of the dataset, scope of behavioral classification and extrapolation to other anonymity protocols. Research in the future ought to focus on multi-protocol detection, fine-grained behavioral classification, temporal analysis, adversarial robustness, and the combination with endpoint security controls.

The privacy protection, free expression, and bypassing censorship are crucial functions of the anonymity networks. At the same time, data exfiltration and other bad practices are also possible due to the same technologies. Our study shows that an organizations networks are able to detect threats without degrading the cryptographic safeguards that render these anonymity networks useful. Both serious and subtle security policy can be achieved through accurate detection and behavioral analysis to defend both the organizational assets and individual privacy rights.

\section*{Declarations}

\subsection*{Funding}
The authors did not receive support from any organization for the submitted work.

\subsection*{Competing Interests}
The authors declare that they have no known competing financial interests or personal relationships that could have appeared to influence the work reported in this paper.

\subsection*{Ethics Approval}
Not applicable. This study used publicly available network traffic data and did not involve human participants or animals.

\subsection*{Data Availability}
The SafeSurf Darknet 2025 dataset used in this study is publicly available at Mendeley Data (DOI: https://doi.org/10.17632/kcrnj6z4rm.2). The trained models and analysis code are available from the corresponding author upon reasonable request.

\subsection*{Code Availability}
Analysis code is available from the corresponding author upon reasonable request.

\subsection*{Author Contributions}
Siddique Abubakr Muntaka: Conceptualization, Methodology, Data collection and curation, Formal analysis, Software implementation, Visualization, Writing - original draft. Muntaka Mohammed and Mansuru Mikail Azindo: Data collection and curation, Formal analysis. Ibrahim Tanko, Franco Osei-Wusu, Edward Danso Ansong, and Benjamin Yankson: Software implementation, Validation. Oliver Kornyo, Foster Yeboah, Jones Yeboah, Richmond Adams, and Pulcheria Serwaa: Visualization, Writing - review and editing. All authors read and approved the final manuscript.

\end{document}